# Demonstration of a novel focusing small-angle neutron scattering instrument equipped with axisymmetric mirrors

Dazhi Liu[1], Boris Khaykovich[1,a], Mikhail V Gubarev[2], J Lee Robertson[3], Lowell Crow[3], Brian D. Ramsey[2], and David E Moncton[1,4]

[1]Nuclear Reactor Laboratory, Massachusetts Institute of Technology, 77 Massachusetts Ave., Cambridge, MA 02139, USA

[2]Marshall Space Flight Center, NASA, VP62, Huntsville, AL 35812 USA

[3]Instrument and Source Design Division, Oak Ridge National Laboratory, Oak Ridge, TN 37831, USA

[4]Department of Physics, Massachusetts Institute of Technology, 77 Massachusetts Ave., Cambridge, MA 02139, USA

**Abstract**

Small-angle neutron scattering (SANS) is the most significant neutron technique in terms of impact on science and engineering. However, the basic design of SANS facilities has not changed since the technique's inception about 40 years ago, as all SANS instruments, save a few, are still designed as pinhole cameras. Here we demonstrate a novel concept for a SANS instrument, based on axisymmetric focusing mirrors. We build and test a small prototype, which shows a performance comparable to that of conventional large SANS facilities. By using a detector with 50-micron pixels, we build the most compact SANS instrument in the world. This work, together with the recent demonstration that such mirrors could increase the signal rate at least 50-fold, while improving resolution, paves the way to novel SANS instruments, thus affecting a broad community of scientists and engineers.

---

[a] E-mail address: bkh@mit.edu

Small-angle neutron scattering (SANS) instruments are found at nearly every neutron-research facility, where they serve a sizable and diverse community of users. The basic instrument design remains the same since original small-angle-scattering instruments were described: a very small aperture, relative to the source size, is used to collimate the beam illuminating the sample.[1] Such a design is not only very inefficient in terms of the neutron flux, it also requires complex and costly detectors, which should move inside vacuum tanks, reaching 10 m in length and more than a meter in diameter. This inefficient design is still in use after decades of developments, demonstrating the enormous challenge of building efficient optics for thermal and cold neutrons.[2] Among notable proposals are the use of zone plates,[3] material and magnetic lenses,[4-6] converging collimators,[7] multiple beams,[8] and mirrors.[9-11] While converging collimators and multiple beams are incremental improvements of the basic design, the use of focusing devices, such as lenses and mirrors, result in qualitatively different instruments, reminiscent of optical microscopes, promising to achieve higher resolution and signal rates. However, focusing devices are used very infrequently because neutron lenses and zone plates have strong chromatic aberrations and focal lengths approaching 100 m, while existing focusing mirrors could not collect a large enough portion of the neutron beam's phase space.

To address this challenge, we pioneered the use of axisymmetric focusing mirrors, called Wolter mirrors, which play the role of achromatic lenses with a short focal length and high throughput.[12, 13] Our ray-tracing analysis of Wolter optics showed the possibility of improving the signal rate by a factor of 50 or more, when optimized for an existing SANS facility.[10] This is a very significant gain, especially if compared with the ten-fold increase of the neutron flux at the new Spallation Neutron Source (SNS) in the US compared with a much older similar facility, ISIS, in the UK. In addition to the gain in the signal rate, the mirror-based instrument could reach lower momentum transfer, than most existing facilities.[10] To demonstrate this novel concept, in this work we have built a prototype Wolter-mirrors-based SANS instrument at Oak Ridge National Laboratory (ORNL). The performance of this instrument, as described below, is in accord with computer simulations and is comparable to that of conventional facilities, even though it is equipped with very small prototype optics. Much larger mirrors can be manufactured for high-performance future facilities. Our demonstration, together with the earlier ray-tracing analysis [10] opens the way for changing SANS instruments from pinhole cameras into microscopes, bringing transformative improvements to rate-limited neutron methods and enabling new science.

## Results

**Experimental setup.** We tested the prototype SANS instrument at an instrument development beamline at ORNL.[14] A schematic layout of the experimental setup is shown in Figure 1. The mirrors play the role of lenses, creating an image of the source aperture at the detector.[13] The mirrors used in this experiment consist of sections of confocal ellipsoid and hyperboloid. Incident rays must reflect from inside surfaces of the optics (ellipsoid, followed by the hyperboloid) before coming to focus. The rays, which do not intersect the first mirror, are stopped by a beamstop in front of the mirrors. The optics used in this experiment is made of three nested co-axial confocal Ni mirrors to improve the throughput. The device, fabricated at NASA using an electroformed replication process, has been used in previous demonstrations of neutron focusing and a neutron microscope.[13,15] For additional drawings and photographs of the set-up, see Supplementary Figs. S1 – S2.

The source aperture and the detector were at two foci of the optics (2.56 m upstream and 0.64 m downstream, respectively). Test samples were placed 0.08 m in front of the detector. The measurements were made in the time-of-flight (TOF) mode, using a neutron chopper installed 0.3 m upstream of the source aperture. The chopper has an opening of 2 degrees and frequency of 40 Hz. The detector, a 50 mm-diameter micro-channel plate with 48 μm pixels, was triggered by the chopper and recorded the position and the TOF of each detected neutron. First, the focal spot and the spectrum of reflected neutrons were measured without the sample. Neutrons from the moderator obey the Maxwell-Boltzmann distribution, which peaks at about 4 Å. The spectrum of the neutrons illuminating the sample is modified by the reflectivity of the mirrors. The spectrum measured at the detector starts from 5 Å and peaks at about 6.5 Å, in accord with expectations from ray tracing.

**Small-angle scattering by test samples.** Two standard test samples, porous silica (Porasil B) and silver behenate powder were used to calibrate neutron intensity and the momentum transfer $Q$, to measure the instrumental resolution and evaluate the low-$Q$ limit of the prototype SANS.

Figure 2a displays experimental results for Porasil B, which is typically used for intensity calibrations of SANS instruments.[16] The scattering intensity from random interfaces in Porasil B is modeled by the Debye-Bueche formula: $I(Q) = I(0)(1 + \alpha^2 Q^2)^{-2}$, where the correlation length $\alpha = 43.5 \pm 2$ Å. Figure 3b shows the Debye-Bueche plot ($I^{-1/2}$ vs $Q^2$) of the data. The linear fit in the range 0.0004 Å$^{-2}$ < $Q^2$ < 0.0012 Å$^{-2}$ gives $\alpha = 38 \pm 4$ Å, consistent with the nominal value. The plot shows that our prototype SANS instrument performs well at $Q$'s down to about 0.02 Å$^{-1}$.

The instrumental resolution is measured utilizing silver behenate, an organic compound, which has ordered lamellar structure. Its first Bragg peak at $Q_0 = 0.1076$ Å$^{-1}$ is commonly used to calibrate momentum transfer measurements.[17] The measured scattering pattern of silver behenate is shown in Figure 3a, where the bright ring is due to the Bragg peak. The figure shows only neutrons from a narrow TOF band, corresponding to the wavelength $\lambda$ = 6.64 Å ± 2%. Figure 3b shows the same peak, integrated over polar angles and wavelengths in the range from 6.4 to 9.2 Å. The width of the Bragg peak gives the $Q$ resolution of the instrument: $\Delta Q/Q_0(\lambda^{-1}) \approx 14\%$, where $\Delta Q$ is the FWHM of the Gaussian peak.

The momentum transfer at small angles $Q = (4\pi/\lambda)\sin(\theta/2) \cong 2\pi\theta/\lambda$, leading to the resolution $\Delta Q/Q_0 = \lambda^{-1} \Delta\theta(\lambda) 2\pi/Q_0$. $\Delta\theta(\lambda)$ is extracted at different wavelengths from the time-of-flight measurements. The instrumental resolution as a function of the inverse neutron wavelength, $\Delta Q/Q_0(\lambda^{-1})$, is plotted in Figure 4. This plot demonstrates that the resolution is determined by the size of the direct beam on the detector. Indeed, the direct beam is the de-magnified image of the source aperture, which is 4 mm in diameter. The direct beam at the detector has a Gaussian profile with the FWHM about 1.25 mm, while the sample-to-detector distance $SDD \approx 80$ mm, leading to $\Delta\theta \approx 0.016$ and $\Delta Q/Q_0 \approx 0.9\,\lambda^{-1}$, which coincides with the linear fit of the data in Figure 4. Other contributions to the resolution, although insignificant, are discussed below.

**Analysis of the instrument resolution.** The resolution function of SANS instruments has contributions from uncertainties of neutron wavelengths and scattering angles.[18-20] Therefore, the resolution is determined by the propagation of relative standard deviations $(\sigma_Q/Q)^2 = (\sigma_\lambda/\lambda)^2 + (\sigma_\theta/\theta)^2$. Here $\sigma_Q, \sigma_\lambda$ and $\sigma_\theta$ denote standard deviations of the momentum transfer ($Q$), wavelength ($\lambda$) and scattering angle ($\theta$) respectively. The beam at the source aperture has an approximately Gaussian profile. Since the samples, the scattering cross-section, and the optics are axisymmetric, we use a one-dimensional approximation. In this case, the resolution $\Delta Q/Q \approx 2.35\, \sigma_Q/Q$. Figure 5a shows the $Q$-dependences of major contributions to the resolution function.

The uncertainty in the wavelength, $\sigma_\lambda$, is due to the finite opening in the beam chopper. The time of flight (TOF) is recorded for each neutron between the chopper and the detector and used to calculate the wavelength: $\lambda[\text{Å}] = 3.956 \cdot \text{TOF}[\text{ms}]/L[\text{m}]$, where $L$=3.5 m is the chopper-to-detector distance. Hence, $(\sigma_\lambda/\lambda)^2 = (\sigma_{\text{TOF}}/\text{TOF})^2$. The uncertainty of the TOF is due to the width of the chopper opening, which is 2 degrees in this experiment. At 40 Hz rotation frequency, the standard deviation of TOF is 0.04 ms by assuming uniform and constant beam flux. For the wavelength range of this experiment, 6.4 to 9.2 Å, corresponding TOF is from 5.67 to 8.14 ms, the standard deviation of 0.04 ms is less than 1%. Since $\sigma_\lambda$ has no momentum-transfer dependence, it is the horizontal line in Figure 5a.

The uncertainty in the scattering angle has four contributions. In addition to the direct beam size ($\sigma_b$), other contributions include the detector pixel size ($\sigma_p$), figure errors of the mirrors ($\sigma_{\text{FE}}$) and the sample size ($\sigma_d$) ($\sigma$'s are variances of the position distributions of scattered neutrons on the detector). To summarize: $\sigma_\theta^2 = \frac{1}{SDD^2}\left(\sigma_b^2 + \sigma_p^2 + \sigma_{\text{FE}}^2 + \sigma_d^2\right)$. The direct beam on the detector is approximately Gaussian with the FWHM of 1.25 mm, leading to 1.25 mm $\approx$ 2.35$\sigma_b$. Therefore, the standard deviation of the angular resolution $\frac{\sigma_b}{SDD} \approx 6.65 \times 10^{-3}$. This is the largest contribution to the resolution function, decreasing with $Q$, as shown in Figure 5a. The uncertainty due to the pixel size, 48 μm in this experiment, is much smaller than the direct beam size. $\sigma_{\text{FE}}$ is determined by the deviation of the mirrors from the ideal shape. The angular resolution of these mirrors is about 100 μrad, corresponding to the positional uncertainty of less than 100 μm at the detector,[15] again much smaller than the direct beam size. Final contribution to the angular uncertainty, $\sigma_d$, is peculiar to the axisymmetric focusing optics. It arises because the sample behind the optics is illuminated by the ring of reflected neutrons. In this experiment, the ring diameter at the sample position was about 4 mm, while the sample-to-detector distance (SDD) was about 80 mm. Figure 5b illustrates the geometry and notations. It shows two neutrons scattered with the same angle $\theta$, but at diametrically opposite points on the ring, reaching the detector at different radii, $d_1$ and $d_2$. Therefore, neutrons with the same scattering angle form a ring with the thickness $\delta d = |d_1 - d_2| = \left|\sqrt{a^2 + s_1^2 - 2as_1 cos\theta} - \sqrt{a^2 + s_2^2 - 2as_2 cos\theta}\right|$, where $a = a_1 = a_2$. Assuming that neutrons are uniformly distributed across $\delta d$, the standard deviation of positions across the scattering ring is $\sigma_d = \delta d/\sqrt{12}$. Numerical calculations of this effect show that it needs to be considered when $Q > 0.3$ Å$^{-1}$ for the particular geometry of this test (see Figure 5b). In future focusing SANS instruments, the SDD will likely be much larger than 80 mm, and so the effect of the ring size will be negligible even at large $Q$'s.

**Discussion**

Let us consider the performance of this prototype SANS instrument, especially at low-$Q$ and high-$Q$ regions, in the interest of future focusing-mirrors-based SANS facilities.

The low-$Q$ limit is determined by the size of the direct beam. For this experiment, we used the beamstop made of a 5 mm wide block of borated aluminum. It is actually larger than the footprint of the transmitted neutron on the detector (about 1.25 mm in diameter). In principle, when an ideal beamstop is installed, the minimum $Q$ of this prototype can reach about $6.7\times10^{-3}$ Å$^{-1}$ in our experimental configuration. It is at the same level as state-of-the-art conventional SANS instruments. If necessary, the low-$Q$ limit can be further extended by using smaller source aperture or by installing another focusing mirror of smaller de-magnification number.

The high-$Q$ limit is determined by the detector's size. For example, if our prototype were equipped with a 100 mm-diameter detector, the highest $Q$ value can reach about 1 Å$^{-1}$, beyond the required highest $Q$ values for most SANS experiments. Figure 5a shows that the resolution at 1 Å$^{-1}$ will be about 10%, which is still very good for most SANS studies.

An important concern of SANS facilities is the background created by diffraction from mirror surface roughness. Such diffraction results in neutrons reflected at large angles (so-called off-specular reflections),[21] potentially creating extra background for weak SANS signals, a limiting factor especially at high $Q$'s. We measured this background in our set-up, using the fact that off-specular scattering is strongly reduced below the cut-off wavelength corresponding to the critical angle. Therefore, we compared the background below and above the cut-off wavelength (about 5 Å) and found no difference between these two cases. Hence, in our measurements, the background introduced by the mirrors is insignificant. Note also, that the operating mirror-based SANS users facility at FRM-II (Munich) does not suffer from effects of surface diffraction, thus demonstrating that state-of-the-art modern polishing techniques are good enough for this purpose.[22]

Gravity effects were not important for this demonstration experiment, as the mirrors-to-mirrors and sample-to-detector distances were only 0.64 m and 0.08 m correspondingly. Since most of the signal is due to neutrons of wavelengths around 6 Å, the gravitational beam deflection was too small to cause a measurable effect. However, for longer future instruments, gravity should be analyzed and compensated, if needed, with the help of prisms, a standard method used at SANS facilities.[2, 19, 23]

To discuss future practical implementations, let us refer to the optimized mirror, which is capable of increasing the signal rate of EQ-SANS, the SANS instrument at SNS, by a factor of 50 or more, as described in Ref. 10. The optic is an elliptical segment of 0.4 m length and 0.09 m radius. This is larger than the prototype device used here, but manageable and entirely within the realm of the existing technology. (EQ-SANS is 13 m long. For shorter focal lengths, two-reflections mirrors should be utilized as in the ellipsoid-hyperboloid prototype used here.) We should note that the large signal rate increase results mostly from the increase in the solid angle of the beam, which illuminates the sample. Therefore, larger samples are necessary to maximize the gain in the signal rate at high resolution, when samples are placed right behind the mirrors. Still, traditional pinhole SANS cannot take advantage of larger samples to increase the intensity, and smaller samples could be placed closer to the detector, leading to increased intensity at some expense to the resolution because of the shorter sample-to-detector distance. Note also, that both the prototype and the mirror analyzed in Ref. 10 are made of only natural Ni. Supermirror Wolter optics is technologically possible,[24] but it is not necessary for SANS in most cases. Wolter optics for SANS could be designed and made within months for a fraction of a cost of a SANS instrument, using standard ray tracing and proven technology. As such, it offers the potential of improving the performance of SANS facilities by orders of magnitude in terms of the signal rate; improve the resolution, and make shorter instruments with simpler detectors.

**Supplementary Information** is available.

**Acknowledgements.** We acknowledge useful discussions with Dr. D. F. R. Mildner (NIST). Research supported by the U.S. Department of Energy, Office of Basic Energy Sciences, Division of Materials Sciences and Engineering under Awards # DE-FG02-09ER46556 and # DE-FG02-09ER46557. The work at ORNL has been sponsored by the Basic Energy Science (BES) Program, Office of Science, U.S. Department of Energy under contract number DE-AC05-00OR22725 with UT-Battelle, LLC.

**Author contributions.** BK, DL, and MVG designed the experiment; DL, BK, MVG, JLR and LC conducted the experiment; DL and BK analyzed the data; MVG and BR made the mirrors; BK, DL, and DEM wrote the manuscript; BK and DEM supervised the project.

**Conflict of interest statement.** The authors declare no competing financial interests.

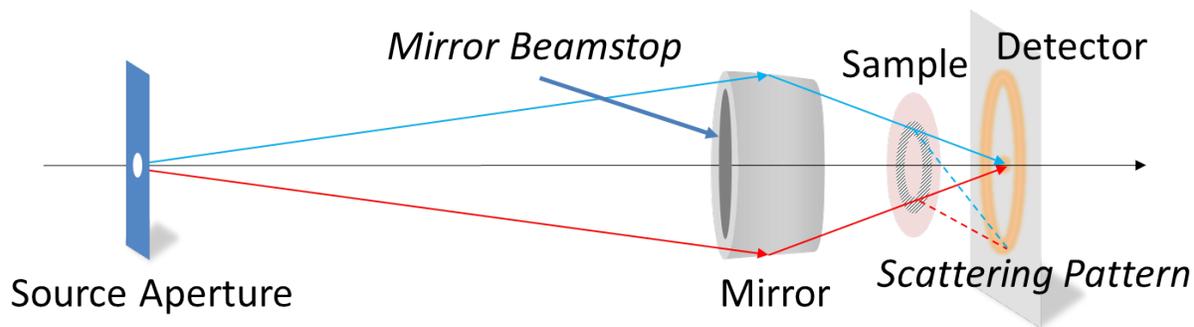

**Figure 1. Schematic layout of the focusing-mirrors-based SANS instrument**. The source aperture and the detector are located at two foci of the mirror. A sample is placed between the optics and the detector. Solid lines indicate trajectories of incident and reflected neutrons, focused at the detector. Dashed lines indicate trajectories of neutrons scattered by the sample. The shaded area of the sample indicates the area, which is illuminated by the beam. The drawing of the mirror is not to scale since the length of the optics is only 60 mm and the distance from the source to the detector is 3.2 m. See Supplementary Figures S1-S3 for additional drawings and pictures of the optics and the setup.

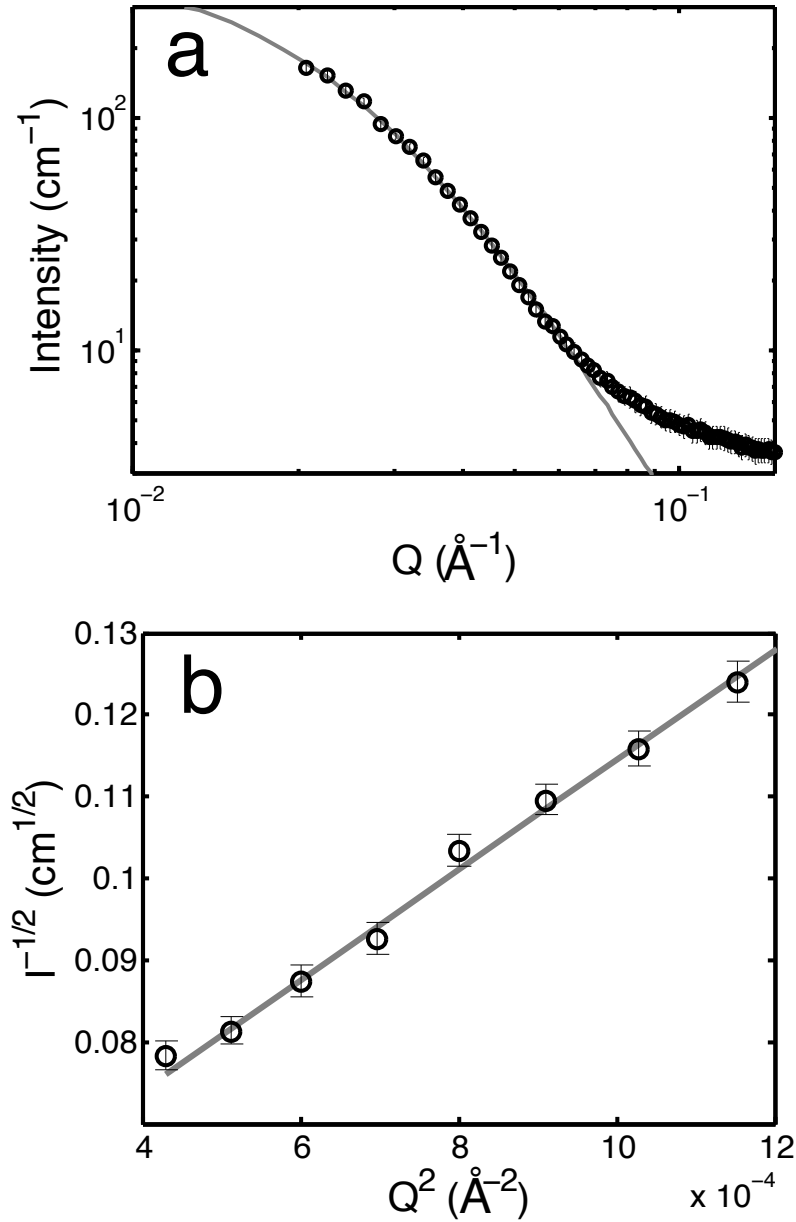

**Figure 2**. **SANS spectra from Porasil B. (a)** The measured SANS curve. Error bars represent standard deviations derived from neutron counts. At low $Q$'s, the error bars are about the size of symbols. **(b)** The Debye-Bueche (D-B) plot of the same data. The D-B plot of the experimental data is linear down to $Q \sim 0.02$ Å$^{-1}$. The dark gray lines (both in (a) and (b)) are the D-B model fit of the data.

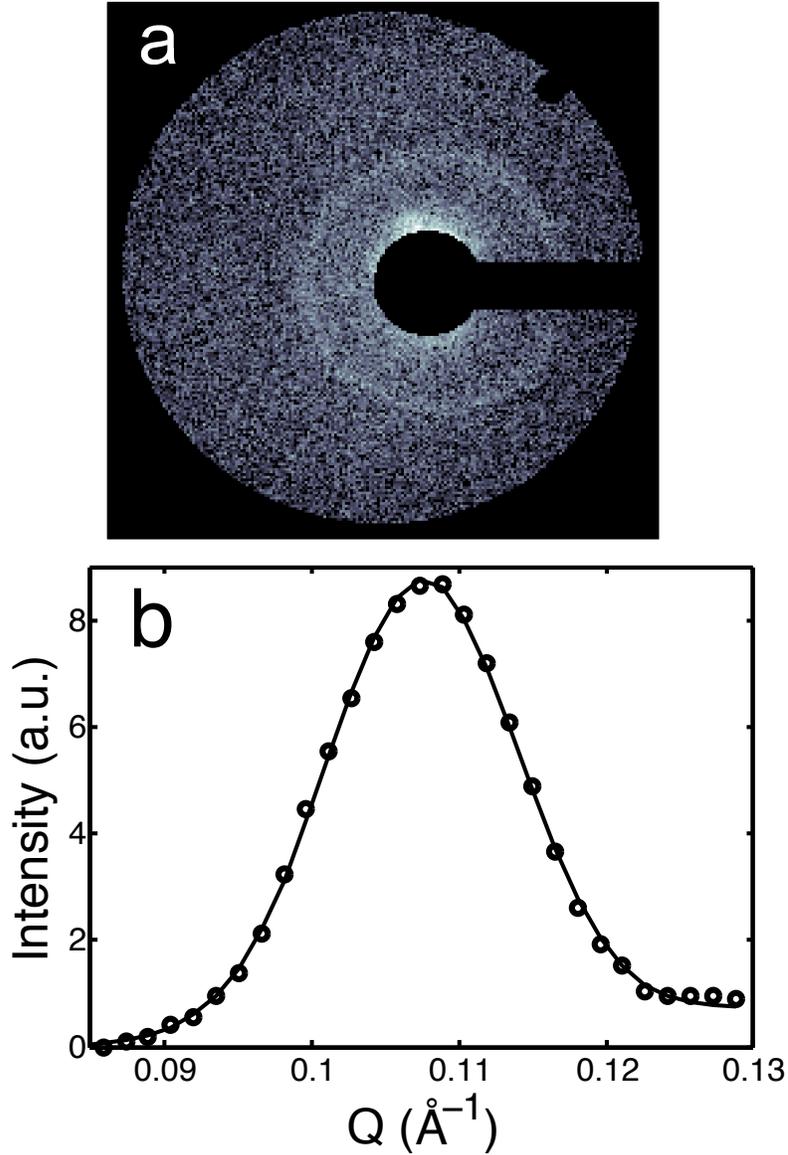

**Figure 3. Neutron diffraction from silver behenate**. **(a)** Raw detector image, using monochromatic neutrons of 6.64 Å ± 2% wavelength (corresponding to the time of flight of 5.88 ms ± 2%). The black area in the center of the image masks the direct beam. The resolution of the detector is 1024×1024 with a 48 micron pitch. The bright ring is the diffraction peak at $Q = 0.1078$ Å$^{-1}$. **(b)** The diffraction peak after integration over the polar angle and wavelengths between 6.4 and 9.2 Å (open symbols). The solid line is a Gaussian fit. The FWHM of the peak ($\Delta Q$) is 0.015 Å$^{-1}$, corresponding to the resolution ($\Delta Q/Q$) of about 14%.

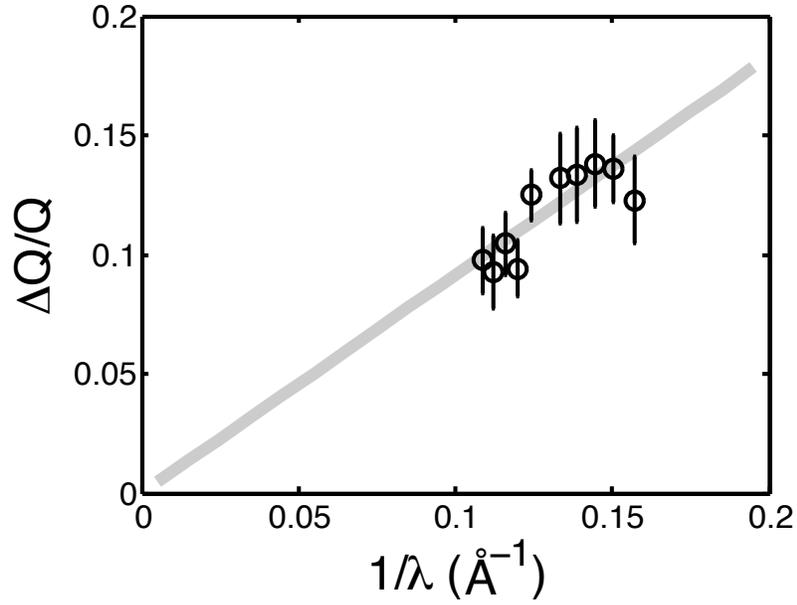

**Figure 4**. **Instrumental resolution as a function of the inverse neutron wavelength**. Open symbols represent $\Delta Q/Q$, where $Q = 0.1078$ Å$^{-1}$ is the positions of the first diffraction peak of silver behenate, and $\Delta Q$ is the FWHM of the peak (see Figure 3b), calculated from the Gaussian fit. Error bars represent 90% confidence interval. The gray line is the resolution function due to the size of the direct beam only (corresponding to $\sigma_b$ in the text). It coincides with the linear fit of the data.

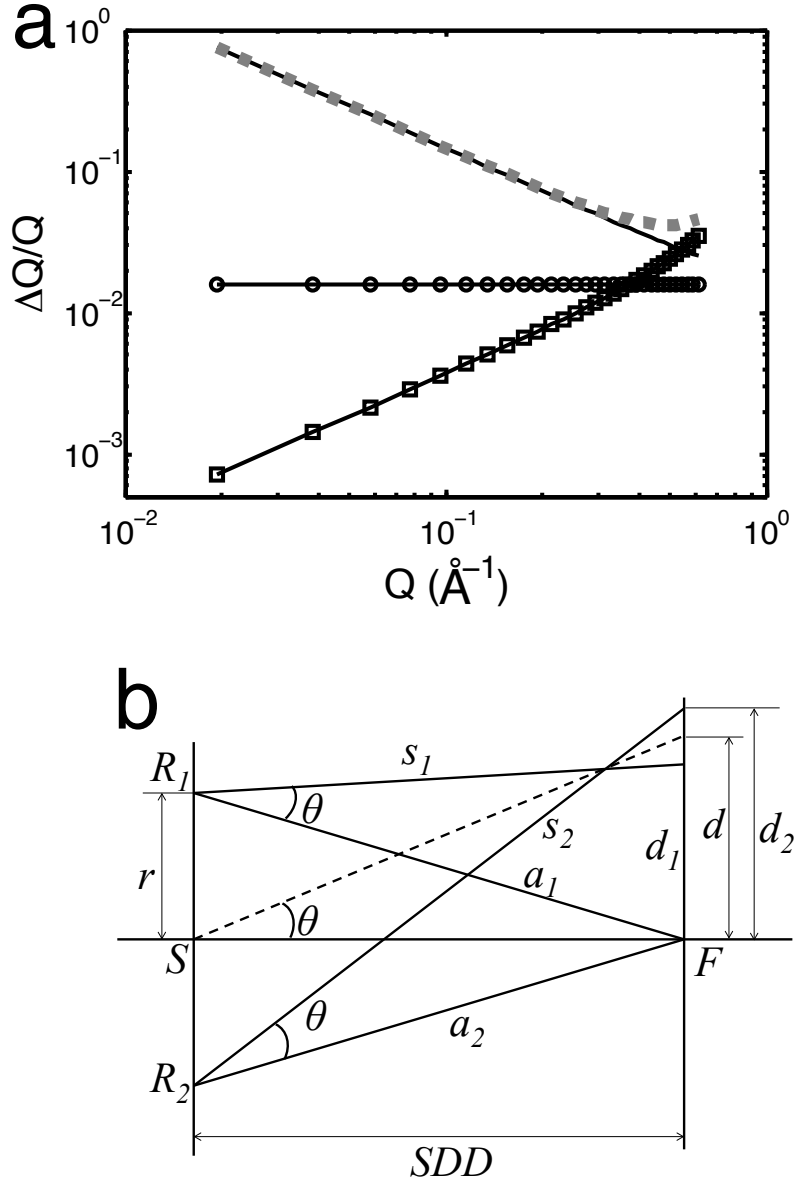

**Figure 5**. **Calculation of the instrumental resolution.** (a) Components of the instrument resolution as a function of the wavector transfer. Squares, circles and triangles represent uncertainties due to the sample size, wavelength spread, and the direct beam size, respectively. The thick gray line shows the overall effect. All the results are calculated at $\lambda = 6.8$ Å. **(b)** The geometry used to calculate the error of $Q$ introduced by the sample size. S and F indicate the sample position (S is the center of the sample ring) and the detector position (F is the focus). Neutrons reflected from axisymmetric optics form a thin ring of radius $r$. For simplicity, the thickness of the ring is not taken into account. $\theta$ is the scattering angle. Lines $a_1$ and $a_2$ are trajectories of the scattered neutrons from points $R_1$ and $R_2$, which are located at the opposite sides of the ring. Line $s_1$ and $s_2$ are the trajectories of scattered neutrons.

**Supplementary Information**

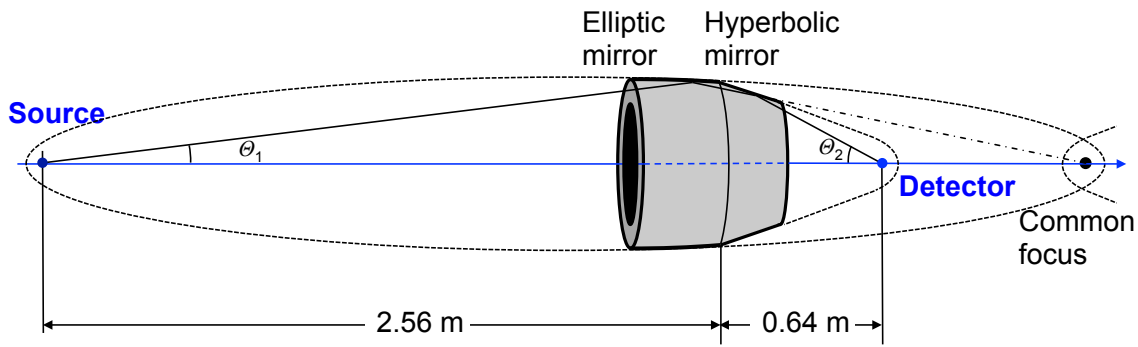

**Supplementary Fig. S1. Schematic drawing of the mirrors used in this test.** The optics is made of segments of confocal ellipsoid and hyperboloid. The source is placed at the focus of the ellipsoid, while the detector is at the focus of the hyperboloid, where the image of the source is created by neutrons, which are reflected from the elliptic and then hyperbolic surfaces. The rest of the beam is stopped by beam-stops and apertures before and after the mirrors. The magnification M = $\Theta_2/\Theta_1$ = 2.56/0.64 = 4. The mirrors are not drawn to scale since their length is only about 60 mm and diameter about 30 mm. Only one mirror is shown for clarity, although three coaxial mirrors of increasing diameters were used (approximately 30, 31.5 and 33 mm), see Reference 13 for details. In the SANS test, the sample is placed between the mirrors and the detector.

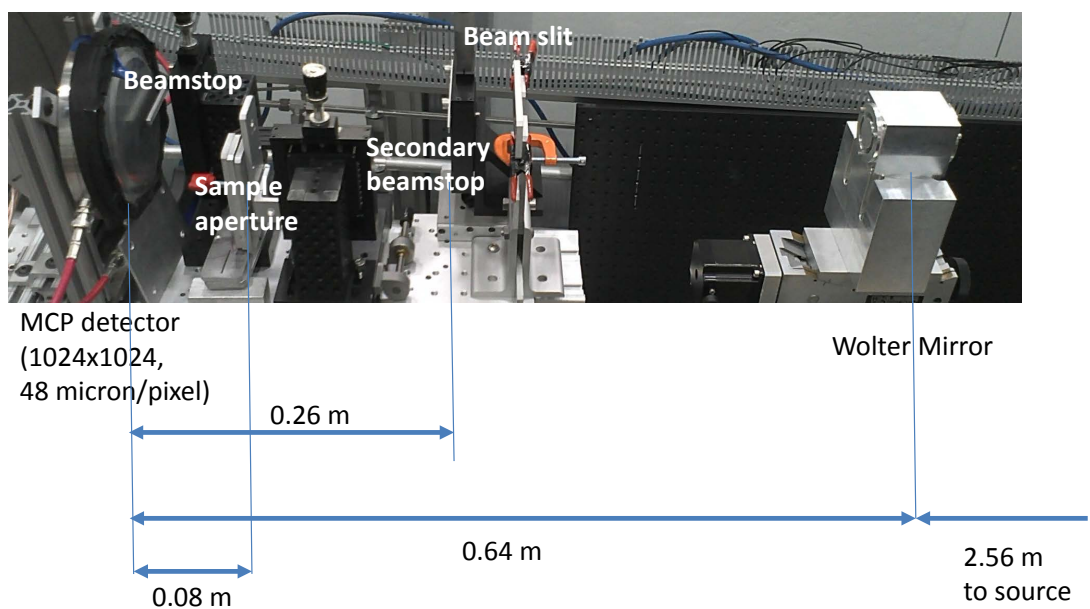

**Supplementary Fig. S2. Photograph of a section of the SANS set-up**. Only the part between the mirrors and the detector is shown, with corresponding distances marked. The detector, the beam stop, the sample aperture, and the mirrors are shown. Additional beam slit and the secondary beamstop were installed to reduce the background from neutrons that were not scattered by both ellipsoid and hyperboloid. The mirrors, the sample aperture and the beam stop were aligned with the help of goniometers and precision stages.